\documentclass[3p]{elsarticle}
\journal{Biographical Memoirs of the Royal Society}
\date{}

\usepackage{amsmath,amssymb,array,booktabs,dsfont,graphicx,mathtools,multirow,rotating,tensor,verbatim}
\usepackage[bookmarksnumbered,linktocpage,pdfstartview=FitH]{hyperref}
\usepackage[all]{hypcap}
\usepackage{graphicx}
\usepackage{caption}
\usepackage{textcomp}
\usepackage{floatrow}

\newcolumntype{M}[1]{>{$}{#1}<{$}}
\newcolumntype{B}[1]{>{\mathbf\bgroup}{#1}<{\egroup}}
\newcolumntype{K}{>{\lvert}{c}<{\rangle}}

\numberwithin{equation}{section}

\newcommand{\be}{\begin{equation}}
\newcommand{\ee}{\end{equation}}
\newcommand{\bea}{\begin{eqnarray}}
\newcommand{\eea}{\end{eqnarray}}

\def\ie{{\it i.e.\ }}

\def\nn{\nonumber} 

\begin{document}

\begin{frontmatter}
\title{
SIR THOMAS WALTER BANNERMAN KIBBLE CBE
\\
~
\\
23rd December 1932 - 2nd June 2016 
\\
~
\\
FRS 1980}

\author[Imperial]{M.~J.~Duff}
\ead{m.duff@imperial.ac.uk}
\author[Imperial]{K.~S.~Stelle}
\ead{k.stelle@imperial.ac.uk}
\address[Imperial]{The Blackett Laboratory, Imperial College London, Prince Consort Road,\\London SW7 2BW, U.K.}
\begin{abstract}
Professor Tom Kibble was an internationally-renowned theoretical physicist whose contributions to theoretical physics range from the theory of elementary particles to modern early-universe cosmology. The unifying theme behind all his work is the theory of non-abelian gauge theories,  the Yang-Mills extension of electromagnetism.
One of Kibble's most important pieces of work in this area was his study of  the symmetry-breaking mechanism whereby the force-carrying vector particles in the theory can acquire a mass accompanied by the appearance of a massive scalar boson.  This idea, put forward independently by Brout and Englert, by Higgs, and by Guralnik, Hagen and Kibble in 1964, and generalised by Kibble in 1967, lies at the heart of the Standard Model and all modern unified theories of fundamental particles. It was vindicated in 2012 by the discovery of the Higgs boson at CERN. According to Nobel Laureate Steven Weinberg, ``Tom Kibble showed us why light is massless''; this is the fundamental basis of electromagnetism.
\end{abstract}
\end{frontmatter}
\begin{figure*}[ht]
\centering
\includegraphics[scale=0.35]{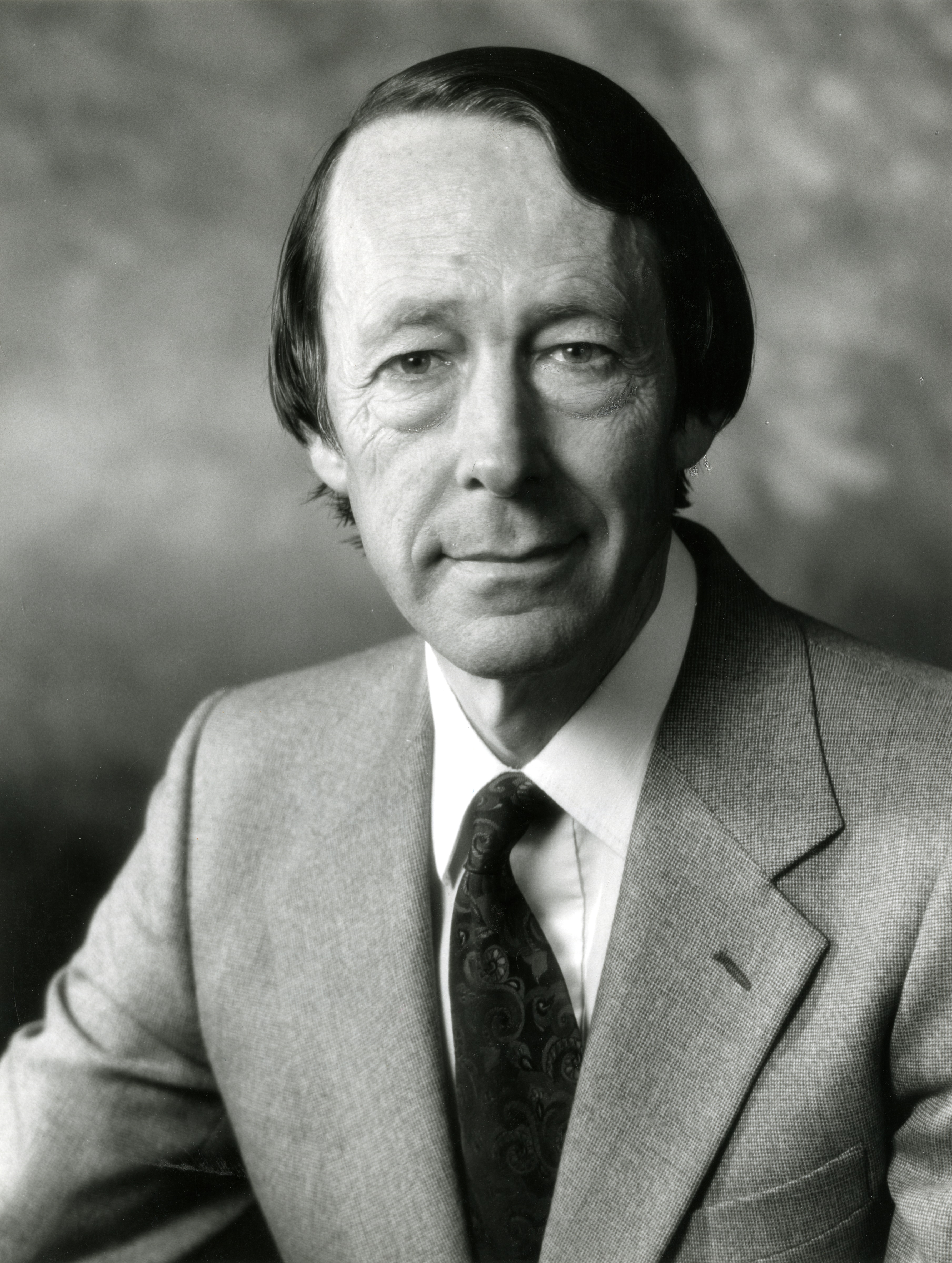}
\caption{Tom Kibble. Portrait by Godfrey Argent 1980 (\textcopyright\  Godfrey Argent Studio)} 
\label{KibbleIC}                     
\end{figure*}

\bigskip
\bigskip
\bigskip
\bigskip

\newpage

\section{\bf EARLY YEARS}
\label{sec:Introductio}
\subsection{\bf India}
\begin{figure*}[ht]
\centering
\includegraphics[scale=1.50]{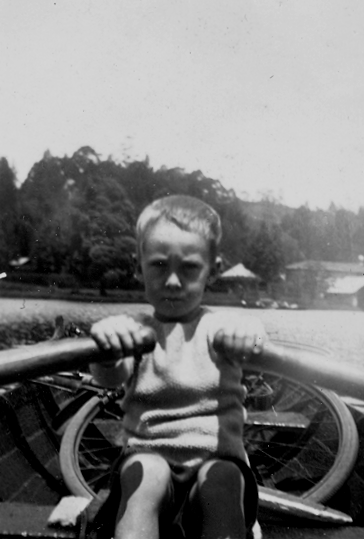}
\caption{Tom rowing 1939 (courtesy of the Kibble family)}
\label{Tom rowing 1939}                     
\end{figure*}

Sir Thomas Walter Bannerman Kibble (FRS 1980), whose portrait is shown in Fig, \ref{KibbleIC}, was born on the 23rd of December 1932 in Madras (now
Chennai), Madras Presidency,
India.
Both his parents were from
missionary backgrounds. His
father Fred was a Professor of
Mathematics at Madras Christian
College.
His mother, Janet Bannerman,
descended from a long line of
ministers in the Church of
Scotland. Her father, William
Burney Bannerman, was a
doctor in the Indian Medical
Service. 
Tom spent his first ten years in
India. He was educated in the
Doveton Corrie School, rated as
one of the best schools in
Madras.
In the summers, his family
generally went to Kodaikanal in
the Palni Hills to escape the heat.
He particularly enjoyed rowing on
the lake there. See Fig. \ref{Tom rowing 1939}.

The principal mathematical influence on young Tom was
his father, who was particularly interested in the symmetry
patterns of Moghul architecture in Agra and elsewhere in
India. 
From this exposure, he gained a love of geometry. His
father taught him about the regular and semi-regular solids,
and he enjoyed making paper models of them.

It had always been intended that at about the age of ten, Tom should
return to Britain for later schooling.
However, this was during the Second World War, and travel was difficult.
In early 1943, passage became available at just two days' notice - he
was not even able to say goodbye to his friends. He went on this voyage
alone.
Since the Suez Canal was closed, the voyage was around Africa, then
across the Atlantic and back in convoy, taking eleven weeks.
In Capetown, Tom went ashore alone at one point but upon returning to
the ship found that it was gone.
Happily, it had simply moved to another dock. Such experiences gave
young Tom a self-reliance which remained with him for life.

\begin{figure}[ht]
\centering
\includegraphics[scale=0.75]{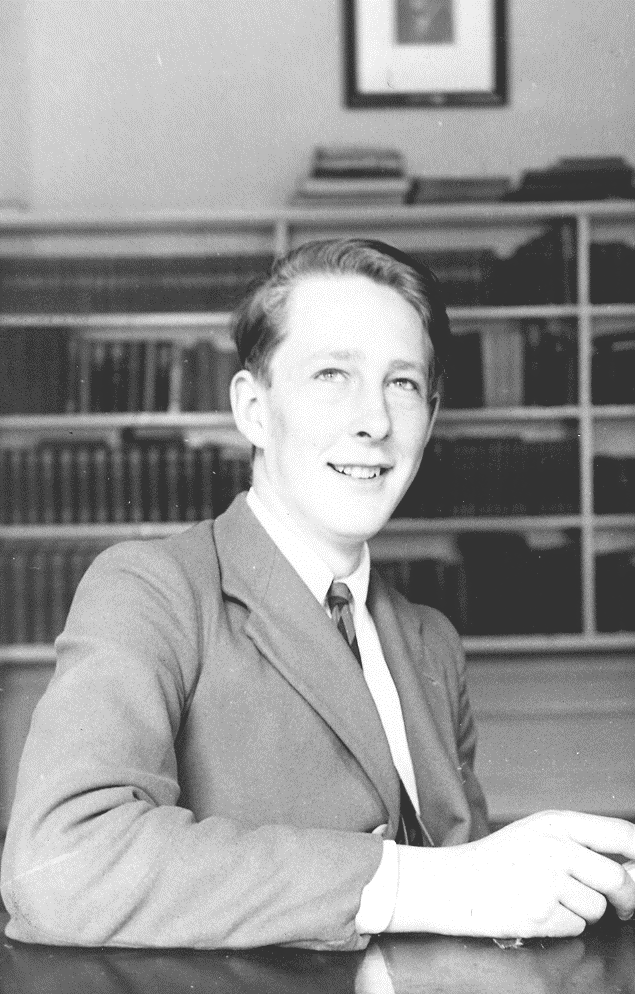}
\caption{Tom Head Boy 1951 (courtesy of the Kibble family)}
\label{Tom Head Boy 1951}                     
\end{figure}

\subsection{\bf Edinburgh}
From 1944 to 1951, Tom attended Melville
College, Edinburgh, becoming Head Boy in
1950-51. See Fig. \ref{Tom Head Boy 1951}.
In his final years at school, he became very
interested in cars, and, together with friends
owned a series of old wrecks.
His father was keen for him to go to Cambridge.
However, in his recollections he said that he
found the problems for the scholarship exams
extremely tedious (and he added that by then
he was enjoying life too much to work on them),
so he was not successful.
Consequently, he went to Edinburgh University,
a decision that he never regretted. 

Two important influences on Tom's future career while an
undergraduate in Edinburgh were Robin Schlapp and
Andrew Nisbet, who taught classical mathematical physics.
Also important was the teaching in pure mathematics in the
Mathematics Department under Alexander Aitken (FRS 1936), who was known
for his great facility for mental arithmetic, being able to
extract cube roots of twelve-digit numbers in seconds.
Tom remarked that it was no accident that a considerable
number of the leading mathematical physicists in the UK
were educated in Edinburgh during this period.
Two other near contemporaries in Edinburgh were David
Olive (FRS 1987) and Keith Moffat (FRS 1986).

Halfway through Tom's undergraduate course, Max Born (FRS 1939)
retired and was replaced by Nicholas Kemmer (FRS 1956), who gave
inspiring lectures completely without notes (but with the
distracting habit of balancing a row of pieces of chalk
end on end).
Tom continued on to a PhD in Edinburgh; his supervisor
was John Polkinghorne (FRS 1974).
His PhD thesis consisted of two parts:
1. Schwinger's Action Principle
2. Dispersion Relations for Inelastic Scattering Processes.

\begin{figure*}[ht]
\centering
\includegraphics[scale=0.75]{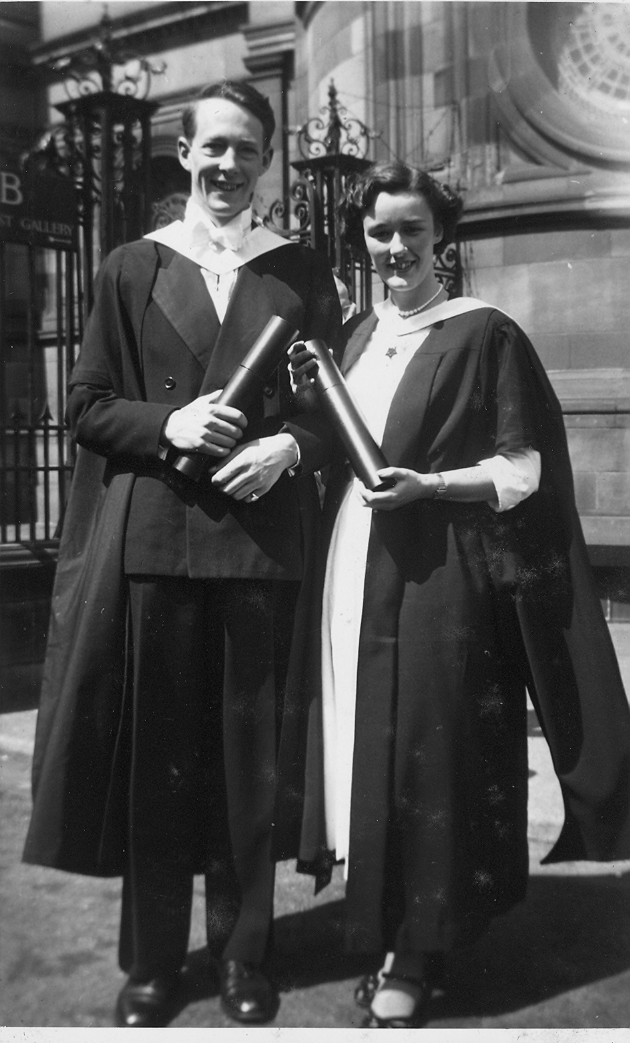}  
\caption{Tom and Anne Graduation (courtesy of the Kibble family)}
\label{TomAndAnneGraduation}                     
\end{figure*}

During Tom's education in
Edinburgh he was for some
time a member of the
Student's Representative
Council, and it was through
this that he met Anne Allan,
who was also a member. See Fig. \ref{TomAndAnneGraduation}.
They became engaged in
June 1954 and married when
he was in the middle of his
PhD, in July 1957.
They had three children:
Helen, Alison and Robert.

\subsection{\bf Caltech}
Following the award of his PhD, Tom received a Harkness
Fellowship to spend a year in the United States. Tom and
Anne went to Caltech, where Tom got to know Dick
Feynman (ForMemRS 1965) and Murray Gell-Mann (ForMemRS 1978).
Upon returning to the UK in 1959 as a NATO Fellow, Tom
joined the Theoretical Physics Group at Imperial College.

\section{\bf IMPERIAL COLLEGE LONDON}
\subsection{\bf Quantum Field Theory and the influence of Abdus Salam}

In 1957 the Head of the Physics Department at Imperial, P. M. S. Blackett (FRS 1933), later Lord Blackett (PRS 1965-1970), recruited the Pakistani physicist Abdus Salam (FRS 1959) from Cambridge to be founding member of The Theoretical  High Energy Physics Group. Salam remained as Professor there until his death in 1996, sharing the 1979 Nobel Prize with Sheldon Glashow and Steven Weinberg (ForMemRS 1981) for unifying the weak and  electromagnetic forces.  As Tom himself recalls [Kibble 1997] ``It was a very exciting place to be. We had lots of visitors: Steven Weinberg, Murray Gell-Mann, Ken Johnson, Lowell Brown, Stanley Mandelstam, John Ward (FRS 1965), to name but a few. The year I arrived,  1959, was also the year Salam became the youngest Fellow of the Royal Society at the age of 33.''

 \begin{figure*}[h]
\centering
\includegraphics[scale=.16]
{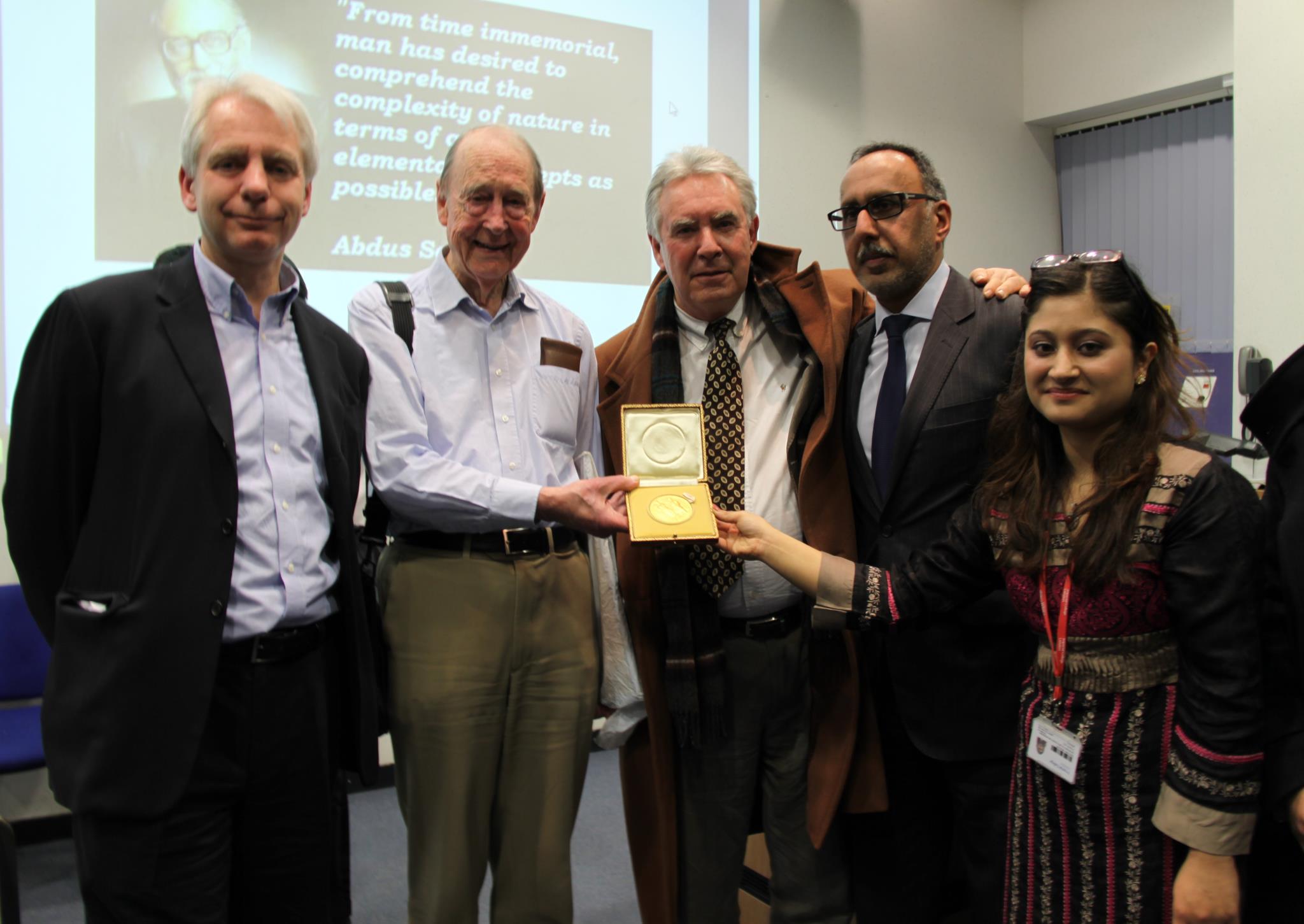}
\caption{Chris Hull, Tom Kibble, Michael Duff, Ahmed Salam, Saba Manzoo with Abdus Salam's Nobel Medal\\ (courtesy of Imperial College)} 
\label{Salammedal}        
\end{figure*}
In the late 50s and early 60s, however, the prevailing mood in fundamental physics 
was not favourable to quantum field theories other than Quantum Electrodynamics (QED).
The success of QED, pioneered by Richard Feynman, Julian Schwinger, Shinichiro Tomanaga and Freeman Dyson (FRS 1952) relied on the smallness of the coupling constant and hence a sensible 
perturbative expansion.This would not be
the case for the strong interactions in the Yukawa theory, however, as
 the squared coupling constant would be about 15. In Tom's own words [Kibble 1997]: ``Salam was convinced from an early date--certainly well before he came to Imperial College---that the fundamental theory would be a gauge theory. This was not a popular view. Many people thought that field theory had had its day, particularly for the strong interaction,  where it was to be superseded by S-matrix theory.  This was an appealingly economical idea: everything was supposed to follow from fundamental principles of covariance, causality and unitarity''.
But Salam was undeterred and under his influence, and that of P. T. Matthews (FRS 1963), the construction of unified
gauge theories was a major theme of the work at
Imperial. Tom Kibble's interest in gauge field theories extended throughout
his career. In 1961, he wrote an influential paper on the relation
between gravity and a gauge theory of the Poincare group [Kibble 1961,Kibble 1986].
Another significant contribution was a series of four papers [Kibble 1968a, Kibble 1968b, Kibble 1968c, Kibble 1968d] devoted to resolving the notoriously tricky problem of defining an S-matrix in theories with massless particles such as QED. There has been a recent revival of interest in this problem and Tom's resolution, involving coherent states, continues to be frequently cited.  See, for example [Hannesdottir 2020] .

  Tom concludes ``I shall always look back on these early years as the most exciting of my life. There was a wonderful atmosphere at the College; I cannot think of anywhere I would rather have been. I shall always be immensely grateful to Abdus Salam for making it possible
 for me to participate in these memorable developments.'' The inspiration was reciprocated: Salam's 1968 electro-weak unification relied crucially on the spontaneous-symmetry-breaking mechanism that Tom pioneered  [Kibble 1964a, Kibble 1967].
 
 \subsection{\bf Spontaneous symmetry breaking}
 
The first example of a gauge theory beyond QED was the SU(2)
non-abelian gauge theory proposed by Yang and Mills [Yang 1954] and
by Salam's student Shaw [Shaw 1955]. Although originally  intended for the strong interactions, this formed the basis of
Salam's unification of the weak and electromagnetic  forces.
One of the 
 key problem for the weak
interactions was their short range, which implies
that the force carrier particles must be massive, as
opposed to the naturally massless excitations of
gauge theories. 

\begin{figure}[ht]
\centering
\includegraphics[scale=.6]
{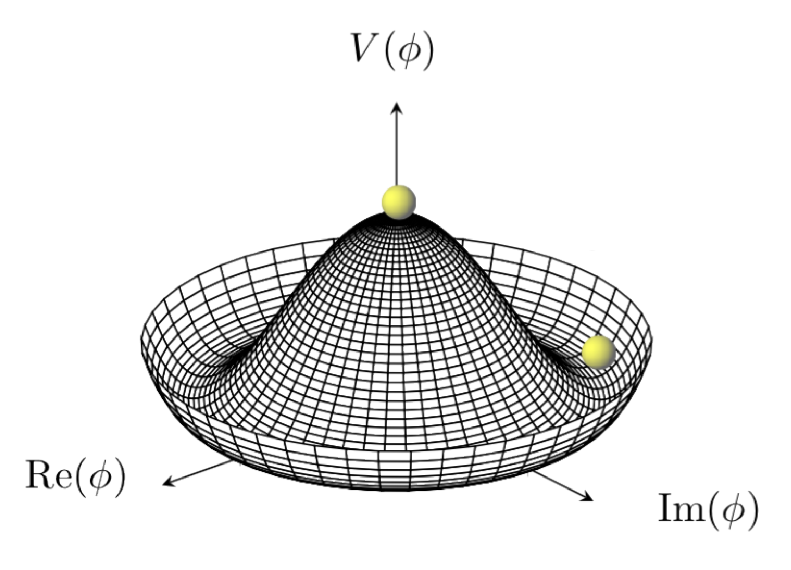}
\caption{Ginzburg-Landau symmetry breaking ``Sombrero'' potential } 
\label{fig:Sombrero_Goldstone}        
\end{figure}

In 1964-65, Gerry Guralnik and Dick Hagen were
visitors. This led to a key collaboration with Tom which borrowed ideas from P.\ Anderson (ForMemRS 1980) in condensed matter physics in the context of the Ginzburg-Landau model of superconductivity. This early model (now derivable from the full Bardeen-Cooper-Schrieffer theory as an effective theory) involves a scalar {\em order parameter} field $\phi(t,{\bf x})$ representing the wave function of a condensate of Cooper pairs, \ie bound state pairs of electrons. The Hamiltonian for the $\phi$ field is
\be
H=\int d^3x\left[\frac1{2m}{\bf D}\phi^\ast\cdot{\bf D}\phi + V(\phi)\right]\nn
\ee
where ${\bf D}\phi=\nabla\phi-2ie{\bf A}\phi$ is the electromagnetic covariant derivative. Expanding in a power series $V(\phi)=\alpha\phi^\ast\phi+\frac12\beta(\phi^\ast\phi)^2$ when $\phi$ is small. Such a potential has a ``sombrero'' shape shown in Fig.\  \ref{fig:Sombrero_Goldstone} if $\alpha<0\,,\ \beta>0$.
The minima of $V(\phi)$ do not occur at the symmetrical point $\phi=0$ but rather at points in the ``rim'' where $\phi^\ast\phi=-\frac\alpha\beta$. This shift to a minimum in the ``rim'' also produces a term $-\frac{2\alpha e^2}{\beta m}{\bf A}\cdot{\bf A}$ giving a mass to the electromagnetic vector field $\bf A$. Consequently, the photon becomes massive in the broken-symmetry phase of a superconductor [Anderson 1963]. 

The idea that a similar mechanism could also give masses to elementary fermions in a relativistic field theory had been suggested by Y.\ Nambu [Nambu 1960]. However, the Ginzburg-Landau model involves an effective scalar-field order parameter which is clearly not a fundamental field of the superconducting theory. The model was also non-relativistic, and the complications imposed by relativity seemed at the time to be crucial. Moreover, giving a mass to the photon was something definitely {\em not} wanted in a fundamental relativistic theory.

 \subsection{\bf Goldstone modes}
 
 Nambu and Jona-Lasinio proposed in 1961  a four-fermion interaction model based on a spinor field $\psi$ with both an ordinary $U(1)$ rigid phase symmetry $\psi\to e^{i\alpha}\psi$ and also a rigid chiral $U(1)$ phase symmetry $\psi\to e^{\tilde\alpha\gamma_5}\psi$ with a Lagrangian interaction term ${\cal L}_{\rm int}=g[(\bar\psi\psi)^2-(\bar\psi\gamma_5\psi)^2]$ [Nambu 1961a, Nambu 1961b]. In this model, a condensate of the $\psi$ field occurs: $\langle\bar\psi\psi\rangle\ne0$. The model successfully gives a mass to the fermion, but it also gives rise to a {\em massless} pseudoscalar mode. Nambu and Jona-Lasinio proposed that, given additional soft explicit chiral symmetry breaking, the pseudoscalar mode could in the end have a small mass and could be identified with the pion. (Indeed, such explicit chiral symmetry breaking is understood now to arise from quark masses in the underlying QCD theory.)
 
 The fact that spontaneous symmetry breaking of rigid symmetries inevitably leads to the existence of massless modes is the content of Goldstone's Theorem, named after Jeffrey Goldstone (FRS 1974) of which a proof was given in 1962 by Goldstone, Salam and Weinberg [Goldstone 1962]. This is illustrated by the two modes of a complex scalar sketched in Fig.\  \ref{fig:Sombrero_Goldstone}. The radial mode clearly experiences a restoring force, but the tangential ``rolling mode'' does not. In a relativistic theory, the radial mode experiencing a restoring force corresponds to a massive particle, while the rolling mode corresponds to a massless particle.
 
 The seeming inevitability of Goldstone modes arising in consequence of symmetry breaking posed a real obstacle to the implementation of the symmetry-breaking idea in relativistic theories, since no such massless bosons were known in physics. Attempts to circumvent the occurrence of such massless Goldstone modes by the addition of explicit symmetry-breaking mass terms were not acceptable when considered at the level of fundamental theory because such terms give rise to non-renormalizability, \ie uncontrollable behaviour at high energies. Such behaviour could be contemplated in an approximate effective theory (such as in pion physics) but not at the level of a fundamental theory of microphysics.

\subsection{\bf The Brout-Englert-Higgs-Guralnik-Hagen-Kibble mechanism}

The resolution of the Goldstone-mode problem came within a short period in 1964 from the work of three independent groups: Brout \& Englert; Higgs; Guralnik, Hagen \& Kibble.

The key point was the recognition that Goldstone's theorem could be evaded in a {\em gauge theory}. Peter Higgs (FRS 1980) had previously shown that, for a gauge theory formulated in radiation gauge, the lack of explicit relativistic Lorentz invariance arising from the gauge choice renders Goldstone's theorem inapplicable.

The three groups approached the symmetry-breaking problem from different perspectives:
\begin{itemize}
\item Brout and Englert, in the first paper to be published, based their argument on a calculation of vacuum polarization in lowest-order quantum perturbation theory about the symmetry-breaking vacuum state [Brout 1964]
\item Higgs's treatment was purely classical and was similar to the Ginzburg-Landau model of superconductivity (which he did not know at the time), but in a relativistic model [Higgs 1964]
\item Guralnik, Hagen and Kibble used a quantum operator formalism, concentrating on the role of the current conservation law and studying  the precise way in which the Goldstone theorem could be evaded [Kibble 1964a]
\end{itemize}
The key conclusion of this fundamental development was that in a gauge theory, the Goldstone mode can be absorbed into the longitudinal mode of the vector gauge field, rendering it massive. The existence of a residual scalar mode, which was discussed only in Higgs's paper, was not considered important at the time.
 
\subsection{\bf Tom Kibble's 1967 paper, the massless photon and the Standard Model}

The three BEHGHK 1964 papers addressed the key problem of how to avoid Goldstone's theorem and generate masses for vector fields by spontaneous symmetry breaking. It was hoped that this could lead to a renormalizable theory with controlled ultraviolet behaviour. (This was later shown by Gerard 't Hooft and Martinus Veltman [Veltman 1972].) However, another key issue remained for a successful application to particle physics: how can one ensure that some gauge fields remain massless, and in particular the photon field?

The three 1964 symmetry-breaking papers had attracted little attention at the time. Meanwhile, in Glashow [Glashow 1961] and Salam and Ward (unaware of Glashow's work) [Salam 1964] had proposed a unified model of the weak and electromagnetic interactions based on the gauge group ${\rm SU}(2)\times{\rm U}(1)$, but without any explanation for the generation of mass needed for the vector particles responsible for the weak interactions.

What was needed to pull these diverse elements together to form what we know know as the Standard Model was a deeper understanding of the general structure of spontaneous symmetry breaking in gauge theories. In 1967, Tom Kibble wrote a magisterial paper [Kibble 1967] laying out the complete mathematical structure of spontaneous symmetry breaking and the Higgs effect. It provided the complete explanation of which vector gauge fields become massive and which remain massless after spontaneous symmetry breaking and the Higgs effect.

This paper was acknowledged by Steven Weinberg in his 1967 paper [Weinberg 1967] as settling the issue the issue of which gauge fields remain massless. This was crucial to the creation of a unified model of the weak and electromagnetic interactions, as the quantum of the electromagnetic field, the photon, had to remain massless. This model, stemming from work by Sheldon Glashow [Glashow 1961] and culminating in the 1967-68 papers of Weinberg and Salam [Salam 1968], led to the award of the Nobel Prize for Physics in 1979. See Fig. \ref{Salammedal}.

A major part of what is now known as the Standard Model, this unification of weak and electromagnetic interactions is based upon the electroweak gauge group ${\rm SU}(2)\times{\rm U}(1)$, which has four Lie algebra generators. The corresponding four vector gauge fields couple to a Higgs complex scalar field gauge-symmetry doublet with a Lagrangian potential term similar to the ``sombrero'' potential sketched in Fig.\ \ref{fig:Sombrero_Goldstone}. The Higgs complex doublet contains four real fields. Three of these belong to the broken-generator coset $({\rm SU}(2)\times{\rm U}(1))/{\rm U}(1)_{\rm em}$, where the unbroken electromagnetic subgroup ${\rm U}(1)_{\rm em}$ mixes one combination of ${\rm SU}(2)$ generators with the ${\rm U}(1)$ ``hypercharge'' factor of the original gauge group prior to symmetry breaking. These erstwhile Goldstone fields are absorbed by the Higgs effect into masses for three of the vector gauge fields. The gauge field for the fourth gauge group generator, corresponding to the unbroken ${\rm U}(1)_{\rm em}$ subgroup, remains massless. This is the electromagnetic photon field $A_\mu$.

To date, the electroweak Standard Model is one of the most precisely tested and verified achievements of elementary particle physics.

\subsection{\bf Cosmic strings}

In 1976, Tom once more wrote an extremely influential
paper on the consequences of spontaneous symmetry
breaking, entitled  `The Topology of Cosmic Domains and Strings' [Kibble 1976]. 
The focus now was on the topological
structure of solutions and domain formation during
phase transitions.

When applied to proposed extensions of the unification idea to include
also the strong interactions, spontaneous symmetry breaking leads to nonsingular magnetic ``monopole'' solutions, as found by  Gerard `t  Hooft ['t Hooft 1974] and
Alexander Polyakov [Polyakov 1974] in an $SU(2)$ model with a triplet Higgs field in
1974. These involve nontrivial maps from the 2-sphere
at spatial infinity to the vacuum manifold $SU(2)/U(1)$. 
Tom's 1976 paper considered the cosmological implications of
more general ``topological defects'': topologically stabilised
strings and domain walls as well as monopoles.
These possibilities are classified by the homotopy groups $\pi_k{(\cal M})$
 of the Higgs vacuum manifold, ${\cal M}$, \ie by maps from
an $S^k$ sphere at spatial infinity to $ {\cal M}$.  If $\pi_0({\cal M}) \neq 0$, one can
have domain walls; if $\pi_1({\cal M}) \neq 0$, one can have string
solutions and if $\pi_2({\cal M} )\neq 0$, one can have monopoles.
Kibble realised that these structures could condense as the universe cooled from the hot conditions prevailing in the big-bang, and might therefore have striking effects on the development of large-scale structure in the universe. 

Finite-temperature corrections induce changes to the
effective potential for the Higgs scalar fields, with an
unbroken gauge symmetry $G$ being restored above some
critical temperature $T_c$. Below this temperature, phase
transitions will take place, producing a phase domain
structure separated by such topological defects.
The values of the symmetry breaking order parameter can
be roughly constant in widely
separated domains, but can be
in conflict where domains meet. This causes formation of a
topological defect, in the center
of which the order parameter
rises to its unbroken symmetry
value, with a consequent
concentration of energy.  In the Big Bang origin of the universe, gauge symmetries
would be restored by the initial high temperatures, with a
rapid thermal freeze-out quench as the temperature
decreased below $T_c$.

The implications of $\pi_0({\cal M}) \neq 0$ domain walls are very
bad for cosmology, since they would be very massive
and would cause an impossibly large anisotropy in the
3{\textdegree}K cosmic microwave background [Kibble 1976]. Consequently, the Higgs
vacuum manifold must be connected. This constraint on models of particle physics heralded
a very fertile interaction between cosmology and
fundamental particle theory.  Similarly there is a danger of an over-abundance of monopoles; a problem nowadays addressed by theories of cosmic inflation.

\begin{figure}[ht]
\centering
\includegraphics[scale=.8]
{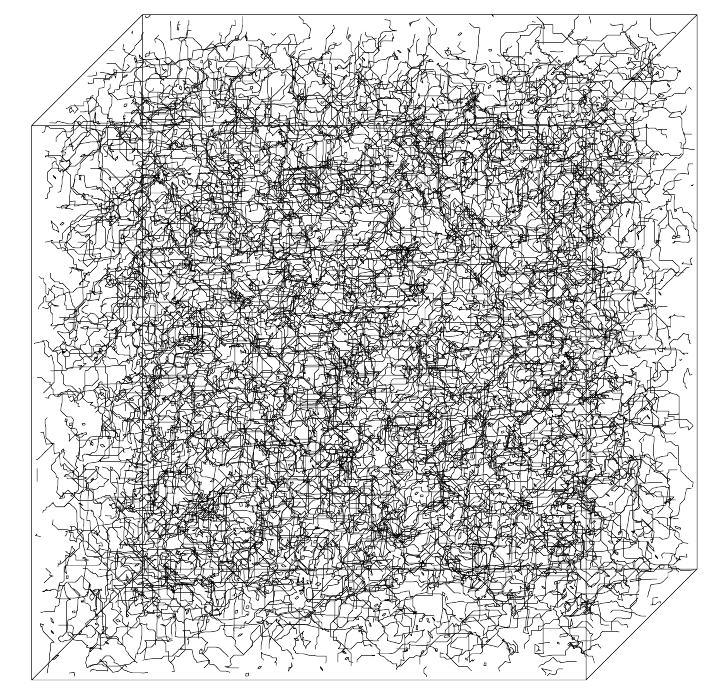}
\caption{Cosmic string network (courtesy of Carlo Contaldi)} 
\label{ffig:cosmicstrings}        
\end{figure}

Cosmic strings, on the other hand, do not suffer
from the same over-dominance
problems as domain walls and
monopoles. Over the years Tom and co-workers including Turok [Kibble 1982] Lazarides [Kibble 1982], Shafi [Kibble 1982], Copeland [Kibble 1993], Hindmarsh [ Kibble 1995], Vilenkin  [Kibble 1995], Davis [Kibble 2005] and Vaschaspati [Kibble 2015] considered models for structure formation seeded by cosmic strings.  Kibble's vision has thereby provided an extraordinary link between the macroscopic and microscopic features of our universe. See Fig. \ref{ffig:cosmicstrings}.

\subsection {\bf String solitons, M-theory and D-branes}

The developments in Kibble's cosmic strings, non-perturbative macroscopic solutions of grand unified quantum field theories, and perturbative superstrings, candidates for the fundamental microscopic constituents of matter incorporating gravity, proceeded along different lines (although in 1985 Witten (ForMem 1999) speculated that fundamental strings produced in the early universe might be stretched to macroscopic scales which however get diluted away by inflation [Witten 1985]).

However, things began to change with the realisation that topological defects such as monopoles, strings  and domain walls appear as  {\it solitonic} solutions of the string equations which, in the low energy limit, are those of supergravity.  Loosely speaking, these solitons are non-singular field configurations which solve the source-free field equations, carry a non-vanishing topological ``magnetic'' charge and are stabilised by a topological conservation law. They are intrinsically non-perturbative with a mass depending on the inverse of the coupling 
constant. See the reviews [Duff 1995] \footnote{ This was co-authored by one of us (MJD) while attending the six-month long workshop
{ \it Topological Defects} organised by Tom Kibble at the Isaac Newton Institute, Cambridge, in 1994.} and [Stelle 1997]. They are to be compared and contrasted with the  {\it elementary} field configurations which are singular solutions of field equation with a source term and carry a non-vanishing Noether ``electric'' charge.

Moreover, since string theory allows ten spacetime dimensions, and its non-perturbative completion M-theory allows eleven, they admit as solutions a menagerie of higher-dimensional objects in addition to the fundamental F-string, such as M-branes, Dirichlet-branes (D-branes) and Neveu-Schwarz branes (NS-branes). They behave like strings if all but one of their spatial dimensions are wrapped around the extra dimensions of spacetime. As Tom Kibble remarked, ``string theory cosmologists have discovered cosmic strings lurking everywhere in the undergrowth''.  In particular Sarangi and Tye [Sarangi 2002] predicted the production of cosmic superstrings in the last stages of brane inflation and Polchinski [Polchinski 2005]  realised that the expanding universe could not only stretch F and D-strings to galactic size but, with large extra dimensions, they could be light enough to survive inflation. Tom continued to make significant contributions to this new cosmic stringy lease of life [Kibble 2010].

Cosmic strings are not the dominant source of density perturbations. Precision measurements of the cosmic microwave background (CMB) suggest that initial random gaussian fluctuations, subsequently amplified by inflation, are the dominant source and 
that cosmic strings cannot account for more than about 10\% of CMB structure. However, some such component could still have observable consequences. Gravitational lensing and gravitational radiation also provide possible observational features of cosmic strings.

\subsection{\bf Condensed matter systems}

Wojciech Zurek pointed out that Tom's description of topological defects also apply to the phase transition of normal fluid helium to superfluid helium [Zurek 1985]. This became known as the``Kibble-Zurek mechanism''. Such a correspondence is found also in other condensed matter systems, for example liquid crystals [Kibble 2002, Kibble 2007].
 Indeed, Tom was involved in an experiment to 
search for defect formation in the B phase of He-3 where the experimental
conditions mimic the cosmological big bang [Kibble 1996].
His work inspired several experimental teams to search for defects 
formed in phase transitions in condensed matter systems, for example Lancaster (He-3 and He-4),  Grenoble and Helsinki (superfluid
systems), US (liquid crystals). These effects have been experimentally confirmed in the context of vortex formation in superfluid Helium 3 [Kibble 1996, Kibble 2002, Kibble 2007].

\section{\bf LEADERSHIP, MENTORING AND SOCIAL RESPONSIBILITY}

  In 1970 Kibble became Professor of Theoretical Physics at Imperial, and held the position of head of the Department of Physics from 1983 to 1991. He was an outstanding teacher and his undergraduate textbook  ``Classical Mechanics '' (now in a 5th edition co-authored with Frank H. Berkshire) is widely regarded as the best of its kind. He was also one of the winners of a competition set by the then Secretary of State William Waldegrave to explain the Higgs Boson to the general public on a single sheet of A4.

  He was elected Fellow of the Royal Society in 1980 and served as its vice-president in 1988-89. At a fairly difficult time, in 1984, he chaired the SERC's Nuclear Physics Board. He received a CBE in 1998 and a Knighthood in 2012. His numerous awards are listed in section {HONOURS AND AWARDS}.

Kibble was concerned about the nuclear arms race and took leading roles in several organisations promoting the social responsibility of science. These included the British Society for Social Responsibility in Science, Scientists Against Nuclear Arms, Scientists for Global Responsibility and the Martin Ryle Trust.

\section{\bf PERSONALITY}
Brilliant physicist,  inspirational teacher and leader though he was, Tom conducted himself with an amazing degree of modesty and humility.  He also showed unfailing kindness to the colleagues and students with whom he came into contact. Tom and his wife Anne frequently invited the  whole Theory Group to parties at their home in Richmond. 

A colleague of Tom at Imperial College, Anne-Christine Davis, now Emeritus Professor of Mathematical Physics at DAMTP Cambridge, writes [Davis 2020]; ``Tom was a mentor to many younger physicists and a strong supporter of women physicists, helping not just those in his immediate vicinity but others in the wider community. At times when one needed a sympathetic ear, or good advice, Tom was there and was at times instrumental in helping female physicists. His dedication to helping and mentoring younger physicists and in particular female physicists was recognised with the 2005 Mentoring lifetime achievement award, NESTA/Nature.  Tom also had a good sense of humour, sometimes dry humour, sometimes even wicked!  At Imperial Tom liked most to be surrounded by others in his research group, in particular the younger members and would make a point of joining them for lunch and tea rather than sitting with senior colleagues.''

 In a moving eulogy [Kibble 2020], his son Robert provided some anecdotes giving us more glimpses of his personality.  One recalls his being introduced to someone who said ``I remember you teaching me thirty years ago, and I've just come back to finish my PhD''  Tom congratulated him, shook his hand, and afterwards asked who that had been, to be told that it was Queen guitarist Brian May.  
Another recalls that while travelling on a bus in his 70s, a pickpocket stole his wallet.  Tom leapt off the bus and chased him down the street, eventually forcing him to drop it. 
Tom was a passionate supporter of many causes, frequently ending up running the organisations he joined. (Scientists Against Nuclear Arms and the Richmond Rambling Association, to name but two.) 
A little known fact is that Tom was an accomplished dancer. 

\section{DISCOVERY OF THE HIGGS BOSON: NOBEL DILEMMA}

\begin{figure}[H]
\centering
\includegraphics[scale=0.25]{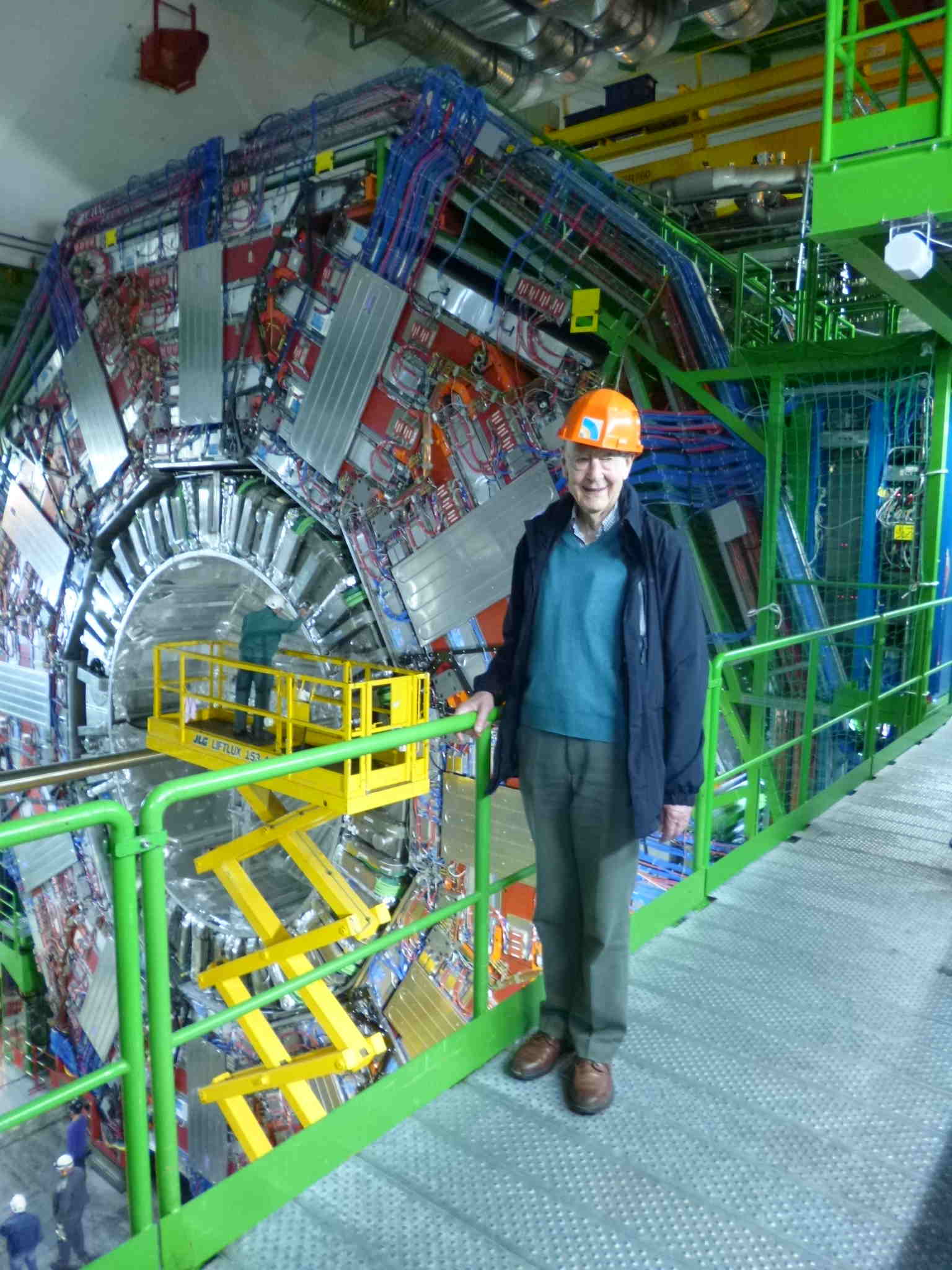}
\caption{Tom Kibble at the CMS Detector, CERN (courtesy of Imperial College)}      
\label{CMS}          
\end{figure}

As we have seen, one of Kibble's most important pieces of work was his study of the spontaneous symmetry-breaking mechanism whereby vector particles can acquire a mass accompanied by the appearance of a massive scalar boson.  This mechanism, put forward independently by Brout and Englert [Brout 1964], by Higgs [Higgs 1964], and by Guralnik, Hagen and Kibble [Kibble 1964] in 1964 lies at the heart of the Standard Model and all modern unified theories of fundamental particles. It was vindicated in 2012 by the discovery of the Higgs boson at CERN; the massive W and Z vector bosons having already been discovered at CERN in 1982 and 1983, respectively. Fig. \ref{CMS} shows Tom standing next to the Compact Muon Solenoid (CMS) which, together with ATLAS, detected the Higgs boson.
This presented a much-debated dilemma for the Nobel Committee, notwithstanding Brout's demise in 2011, as the Prize can be awarded to at most three.  However, Kibble was also sole author of a 1967 paper, focusing on the non-abelian generalisation.  Indeed, in  early papers, the mechanism was known as the ``Higgs-Kibble effect''. See, for example, the 1972 paper by `t Hooft and Veltman [Veltman 1972] for which they won the 1999 Nobel Prize. Higgs and Kibble were the joint winners of the 1981 Hughes medal of the Royal Society and the 1984 Rutherford Medal and Prize of The Institute of Physics.

Tom's 1967 paper laid out the
complete mathematical structure of spontaneous symmetry breaking,  settling the issue of which gauge fields
become massive and which remain massless. Private discussions with Salam informed his thinking on the problem.
For a gauge field theory model based on a non-abelian gauge group $G$ and a scalar Higgs sector producing
spontaneous symmetry breaking down to a subgroup $H$, vector fields corresponding to the coset $G/H$ become
massive, while the vectors corresponding to the unbroken subgroup $H$ remain massless. When the Standard Model came along shortly afterwards, Kibble's paper was seen to explain not only why the W and Z acquire a mass but, equally crucial, why the photon does not. His 80th birthday was marked by Imperial College in March 2013 with a symposium day. In his evening public lecture, Nobel Laureate Steven Weinberg [Weinberg 2013] ended by saying ``Tom Kibble showed us why light is massless''.  Indeed Higgs said that Kibble should have shared the 2013 Nobel Prize awarded to Englert and himself ``because of what he wrote in 1967'' [Higgs 2012a, Higgs 2012b]. Kibble himself maintained a dignified modesty throughout in keeping with the honesty and integrity for which he was justly famous.

\newpage

\section{\bf  HONOURS AND AWARDS}

1980   Fellow of the Royal Society\medskip

1981   Hughes Medal of the Royal Society (with P.W. Higgs)\medskip

1984   Rutherford Medal \& Prize, Institute of Physics (with P.W. Higgs)\medskip

1991   Fellow of the Institute of Physics\medskip

1993   Guthrie Medal \& Prize, Institute of Physics\medskip

1998  CBE\medskip

2005 Mentoring lifetime achievement award, NESTA/Nature\medskip

2009 Dirac Medal IOP\medskip

2010 Sakurai Prize for Theoretical Physics\medskip

2012 Royal Medal, Royal Society\medskip

2012 Honorary Fellow IOP\medskip

2012 Einstein Medal\medskip

2013 Knighthood\medskip

2014 Royal Medal, Royal Society of Edinburgh\medskip

2016 Isaac Newton Medal IOP

\section{\bf REFERENCES TO OTHER   AUTHORS}

\noindent
[Yang 1954] Conservation of isotopic spin and isotopic gauge invariance. Yang, C. N. and Mills, R. L.\\ {\it Phys, Rev., 96 191-195} (doi:10.1103/PhysRev.96.191)\medskip

\noindent
[Shaw 1955] Shaw, R. PhD thesis, Cambridge University (Sept. 1955)\medskip

\noindent
[Nambu 1960] Dynamical theory of elementary particles suggested by superconductivity. Nambu, Y.\\ In {\it Proceedings of ICHEP 60}, 858-866\medskip

\noindent
[Nambu 1961a] Dynamical Model of Elementary Particles Based on an Analogy with Superconductivity, I. Nambu, Y. and Jona-Lasinio, G. {\it Phys.Rev. 122 (1961) 345-358}
 (doi:10.1103/PhysRev.122.345)\medskip
 
\noindent
[Nambu 1961b] Dynamical Model of Elementary Particles Based on an Analogy with Superconductivity, II.  Nambu, Y. and Jona-Lasinio, G. {\it Phys.Rev. 124 (1961) 246-254}
 (doi:10.1103/PhysRev.124.246)\medskip

\noindent
[Glashow 1961] Partial Symmetries of Weak Interactions. Glashow, S.\\ {\it Nucl. Phys. \textbf{22} (1961), 579-588}
(doi:10.1016/0029-5582(61)90469-2)\medskip

\noindent
[Goldstone 1962] Broken symmetries. Goldstone J., Salam, A. and Weinberg, S.\\ {\it  Phys.Rev. 127  965-970} (doi:10.1103/PhysRev.127.965)\medskip

\noindent
[Anderson 1963] Plasmons, gauge invariance, and mass. Anderson, P. W.\\ {\it Phys. Rev. 130 439-442}
 (doi:10.1103/PhysRev.130.439)\medskip

\noindent
[Brout 1964] Broken Symmetry and the Mass of Gauge Vector Mesons. Brout, R. and Englert, F.\\ {\it Phys. Rev. Lett. 13 (1964) 321-323} (doi:10.1103/PhysRevLett.13.321)\medskip

\noindent
[Higgs 1964] Broken symmetries, massless particles and gauge fields. Higgs, P.\\ {\it  Phys.Lett. 12 (1964) 132-133} (doi:10.1016/0031-9163(64)91136-9)\medskip

\noindent
[Salam 1964]  Electromagnetic and weak interactions. Salam, A. and Ward, J.\\ {\it Phys. Lett. 13 (1964),168-171}
(doi10.1016/0031-9163(64)90711-5)\medskip

\noindent
[Weinberg 1967] A Model of Leptons. Weinberg, S.\\ {\it Phys. Rev. Lett. 19 (1967), 1264-1266}
(doi:10.1103/PhysRevLett.19.1264)\medskip

\noindent
[Salam 1968] Weak and electromagnetic interactions. Salam, A. In {\it Elementary Particle Physics: Relativistic Groups and Analyticity, N. Svartholm, Ed., Almquvist and Wiksell, Stockholm (1968) 367, Proceedings of the Eighth Nobel Symposium.}\medskip

\noindent
[Veltman 1972] Regularization and renormalization of gauge fields. 't Hooft, G. and Veltman, M.\\ {\it Nuclear Physics B, 44(1), 189-213} (doi:10.1016/0550-3213(72)90279-9)\medskip

\noindent
['t Hooft 1974] Magnetic monopoles in unified gauge theories. 't Hooft, G.\\ {\it Nucl. Phys. B 79 276-284}
(doi:10.1016/0550-3213(74)90486-6)\medskip

\noindent
[Polyakov 1974]  Particle Spectrum in the Quantum Field Theory. Polyakov, A. M.\\ {\it  JETP Letters. 20 (6): 194-195}
( ISSN 0370-274X)\medskip

\noindent
[Zurek 1985] Cosmological experiments in superfluid helium? Zurek, W.\\ {\it Nature 317 (6037): 505-508} ( doi:10.1038/317505a0)\medskip

\noindent
[Witten 1985] Cosmic superstrings. Witten, E.\\ {\it Phys. Lett. B 153, 243-246} (doi:10.1016/0370-2693(85)90540-4).\medskip

\noindent
[Stelle 1997] BPS branes in supergravity. Stelle, K. S.\\  In {\it Trieste 1997, High energy physics and cosmology 29-127}\medskip

\noindent
[Sarangi 2002] Cosmic string production towards the end of brane inflation.  Sarangi, S. and Tye, S. H.\\ {\it Physics Letters B. 536 (3-4): 185} (doi:10.1016/S0370-2693(02)01824-5. S2CID 14274241)\medskip

\noindent
[Copeland 2004] Cosmic F- and -D strings. Copeland, E., Myers, R. and Polchinski, J.\\ {\it JHEP 06 013-013}
(doi: 10.1088/1126-6708/2004/06/013)\medskip

\noindent
[Polchinski 2005] Cosmic superstrings revisited. Polchinski, J.\\ {\it  Int. J. Mod. Phys. A 20, 3413-3415} (doi:10.1142/S0217751X05026686).\medskip 

\noindent
[Weinberg 2012]
http://www3.imperial.ac.uk/newsandeventspggrp/imperialcollege/eventssummary/event\_21-1-2013-16-35-46\medskip

\noindent
[Higgs 2012a]
({\it The Scotsman}, 13 November 2013)\medskip

\noindent
[Higgs 2012b] 
http://www.telegraph.co.uk/science/science-news/10443449/Peter-Higgs-Tom-Kibble-should-have-shared-Nobel-Prize-with-me.html\medskip

\noindent
[Hannesdottir 2020]  An S-matrix for massless particles.   Hannesdottir, H. and Schwartz, M. D.\\ {\it Phys Rev D101  105001}.
(doi:10.1103/PhysRevD.101.105001)\medskip

\noindent
[Davis 2020] Private Communication. Davis, A-C.\medskip

\noindent
[Kibble 2020] Private Communication. Kibble, R.


\section{BIBLIOGRAPHY}


\noindent
[Kibble 1961]
Lorentz invariance and the gravitational field.\\ {\it  J. Math. Phys. 2, 212-221} (doi: 10.1063/1.1703702)\medskip


\noindent
[Kibble 1964a] 
(with Guralnik, G. S. and Hagen, C. R.) Global Conservation Laws and Massless Particles.  {\it Phys.Rev.Lett. 13 (1964) 585-587} (doi:10.1103/PhysRevLett.13.585)\medskip

\noindent
[Kibble 1964b]
(With Brown, L. S.) Interaction of intense laser beams with electrons.\\ {\it Phys. Rev. 133,
A705-19}
(doi:10.1103/Phys. Rev.133.A)\medskip

\noindent
[Kibble 1967]
Symmetry-breaking in non-Abelian gauge theories.\\ {\it Phys. Rev. 155 (1967), 1554-1561} (doi:10.1103/Phys Rev.155.1554)\medskip

\noindent
[Kibble 1968a]
Coherent Soft-Photon States and Infrared Divergences. i. Classical Currents.\\
{\it J. Math. Phys. 9 no. 2  315-324}
(doi: 10.1063/1.16645820)\medskip

\noindent
[Kibble 1968b]
Coherent Soft-Photon States and Infrared Divergences. ii. Mass-Shell Singularities of Green's Functions.
{\it Phys. Rev. 173 1527-1535}
(doi: 10.1103/PhysRev.173.1527)\medskip

\noindent
[Kibble 1968c]
Coherent Soft-Photon States and Infrared Divergences. iii. Asymptotic States and Reduction Formulas. 
{\it Phys. Rev. 174 1882-1901}
(doi:10.1103/PhysRev.174.1882)\medskip

\noindent
[Kibble 1968d]
Coherent Soft-Photon States and Infrared Divergences. iv. The Scattering Operator.\\ 
{\it Phys. Rev. 175  1624-1640}
(doi: 10.1103/PhysRev.175.1624)\medskip

\noindent
[Kibble 1976]
Topology of cosmic domains and strings.\\ {\it J. Phys. A: Math. \& Gen. 9, 1387}
 (doi: 10.1088/0305-4470/9/8/029)\medskip

\noindent
[Kibble 1980]  Some Implications of a Cosmological Phase Transition.\\ {\it Phys. Rept. 67, 183} (doi: 10.1016/0370-1573(80)90091-5)
\medskip

\noindent
[Kibble 1982]
(with  Turok, N.) Selfintersection of cosmic strings.\\ {\it Phys.Lett.B 116, 141-143} (doi: 10.1016/0370-2693(82)90993-5)\medskip

\noindent
[Kibble 1982]
(With Lazarides, G. and  Shafi, Q.) Walls bounded by strings.\\ {\it Phys. Rev. D26 435-439} (doi: 10.1103/PhysRevD.26.435)\medskip

\noindent
[Kibble 1986]
(With Stelle, K. S.) Gauge theories of gravity and supergravity. In {\it Progress in quantum field
theory, ed. H. Ezawa \& S. Kamefuchi (Amsterdam: North-Holland, 57-81.
Festschrift for H. Umezawa}\medskip

\noindent
[Kibble 1993] 
(with Austin, D. and Copeland, E.) Evolution of cosmic string configurations.\\ {\it Phys. Rev. D48 5594-5627} (doi: 10.1103/PhysRevD.48.5594)\medskip

\noindent
[Kibble 1995]
(With Hindmarsh, M. B.) Cosmic strings.\\ {\it Reports on Progress in Physics 58, 477-562} (doi: 10.1088/0034-4885/58/5/001)\medskip

\noindent [Kibble 1995]
(with Vilenkin, A.)  Phase equilibration in bubble collisions.\\ {\it Phys.Rev.D 52 (1995) 679-688}
(doi:10.1103/PhysRevD.52.679)\medskip

\noindent
[Kibble 1996]
(With Ruutu, V .M.,  Eltsov, V. B., Gill, A. J., Krusius, M., Makhlin, Yu. G., Pla\c{c}ais,  B.,
 Volovik, G. E.  and  Xu Wen), Vortex formation in neutron-irradiated superfluid 3He-B as 
an analogue of cosmological defect formation. {\it Nature 382 (1996) 334-336}
(doi:10.1038/382334a0)\medskip

\noindent
[Kibble 1997]
Recollections of Abdus Salam at Imperial College. In {\it The Abdus Salam Memorial Meeting, eds Ellis, Hussain, Kibble, Thompson, Virasoro, World Scientific 1999} \medskip

\noindent
[Kibble 2002]
Symmetry breaking and defects. In {\it Patterns of Symmetry Breaking, ed. H. Arodz, J.
Dziarmaga \& W.H. Zurek, NATO Science II, 127, pp. 3-36, Proceedings of NATO
Advanced Study Institute, Cracow, September 2002}\medskip

\noindent
[Kibble 2005]
(With  Davis, A-C.)  Fundamental cosmic strings.\\ {\it Contemporary Physics, 46, 313-322}
(doi:10.1080.00107510500165204) \medskip

\noindent
[Kibble 2007]
Phase transition dynamics in the lab and the universe.\\ {\it  Physics Today 60, no. 9, 47-53}
(doi:10.1063/1.2784684) \medskip

\noindent
[Kibble 2009]
Englert-Brout-Higgs-Guralnik-Hagen-Kibble mechanism. {\it Scholarpedia 4(1):644}\medskip

\noindent
[Kibble 2010]
(With Copeland, E. J.) Cosmic strings and superstrings.\\ {\it Proc. Roy Soc. A 466, 623-657}
(doi:10.1098.rspa.2009.0591)\medskip

\noindent
[Kibble 2015]
(With Vachaspati, T.) Monopoles and strings.\\ {\it J. Phys. G42, 094002}
(doi:10.1088/0954-3899/42/9/094002)\medskip

%

\section{\bf ACKNOWLEDGEMENTS}
We thank Carlo Contaldi, Anne Davis, Graziela De Nadai, Philip Diamond, John Ellis, David Fairlie, Chris Isham, Robert Kibble, Alison Martin, Arttu Rajantie, Ray Rivers and Helen Wilson, for help in preparing this memoir.

\section{\bf AUTHOR PROFILES}

\begin{itemize}
\item{\bf Michael Duff}\\

Michael Duff (FRS 2009) is Emeritus Professor of Theoretical Physics and Senior Research Investigator at Imperial College London.  He was formerly Abdus Salam Professor of Theoretical Physics and Principal of the Faculty of Physical Sciences.
His time in the Theory Group overlapped with that of Tom Kibble for a total of 23 years.
He is recipient of the 2004 Meeting Gold Medal, El Colegio Nacional, Mexico, the 2017 Paul Dirac Gold Medal and Prize, Institute of Physics, UK and the 2018 Trotter Prize. He is a visitor to The Institute of Quantum Science and Engineering and was inducted as 2019 Faculty Fellow of the Hagler Institute for Advanced Study, both at Texas A\&M University. 

\item{\bf Kellogg Stelle}\\ 

Kelly Stelle  is  Professor of Theoretical Physics at Imperial College London and formerly Head of the Theoretical Physics Group. He is the recipient of the 2006 Humboldt Research Award, Alexander von Humboldt Foundation, and the 2020 John William Strutt, Lord Rayleigh Medal and Prize, Institute of Physics.
\end{itemize}
  
  \begin{figure*}[h]
    \centering
    \begin{floatrow}\hspace{1cm}
      \ffigbox[6.5cm]{\caption*{Michael Duff FRS (Portrait by Beyond the Pixels.)}}{%
      \includegraphics[scale=0.165]{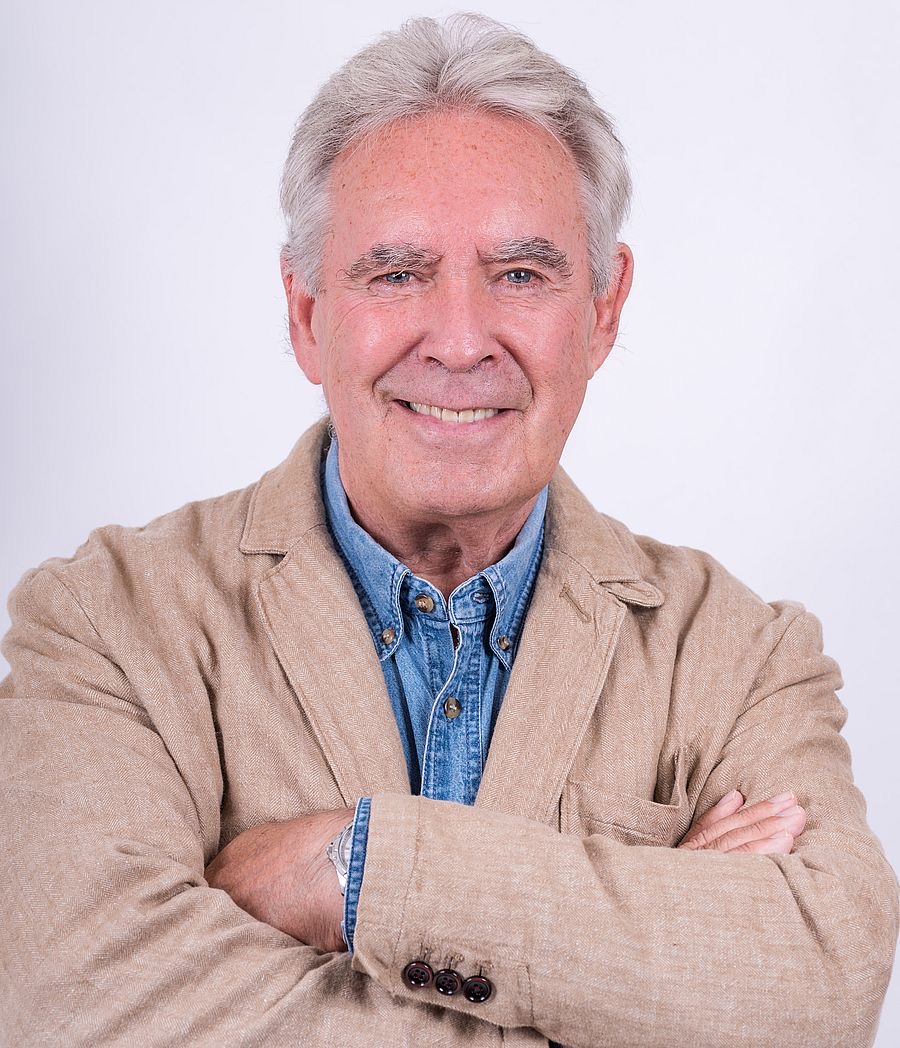} 
      }
      \hspace{1cm}
      \ffigbox[6.5cm]{\caption*{Kellogg Stelle (courtesy of Imperial College)}}{%
      \includegraphics[scale=0.60]{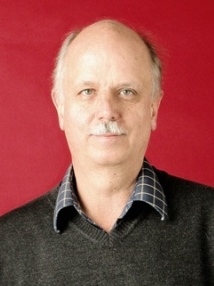}
      }
    \end{floatrow}
  \end{figure*}

\end{document}